# Asymmetric damage avalanche shape in quasi-brittle materials and sub-avalanches (aftershocks) clusters

Chi-Cong Vu[1] and Jérôme Weiss[2,*]

[1]*National University of Civil Engineering, Ha Noi, Viet Nam*
[2]*Univ. Grenoble Alpes, CNRS, ISTerre, 38000 Grenoble, France*

Crackling dynamics is characterized by a release of incoming energy through intermittent avalanches. The shape, i.e. the internal temporal structure of these avalanches, gives insightful information about the physical processes involved. It was experimentally shown recently that progressive damage towards compressive failure of quasi-brittle materials can be mapped onto the universality class of interface depinning when considering scaling relationships between the global characteristics of the microcracking avalanches. Here we show, for three concrete materials and from a detailed analysis of the acoustic emission waveforms generated by microcracking events, that the shape of these damage avalanches is strongly asymmetric, characterized by a very slow decay. This remarkable asymmetry, at odds with mean-field depinning predictions, could be explained, in these quasi-brittle materials, by retardation effects induced by enhanced viscoelastic processes within a fracture process zone generated by the damage avalanche as it progresses. It is associated with clusters of sub-avalanches, or aftershocks, within the main avalanche.

When slowly driven, many physical systems crackle [1], i.e. respond to slowly varying external conditions through intermittent discrete bursts, or avalanches, spanning a wide range of sizes and/or energies [2]. Classical examples are Barkhausen noise in magnets [3], sheared granular media [4-6], plasticity in crystalline [7,8] and amorphous [9] materials, domain walls [10], or martensitic phase transitions [11]. Damage, fracture and friction within heterogeneous materials also release the incoming mechanical energy through intermittent events, from the laboratory scale [12-14] to earthquakes [15]. In these systems, avalanches are characterized by power laws and scaling relationships indicating the emergence of scale invariance and criticality. A key question is whether these various systems belong to a unique universality class [1,16,17], with the potential consequence of making universal predictions from the analysis of a representative model. This has been largely addressed from the analysis of distributions of avalanche sizes, energies and durations, as well as scaling relations between these variables, hence defining a set of critical exponents. In particular, it has been argued that several of these systems could belong to the well-described universality class of interface depinning [18-20], for which mean field [21] and beyond mean-field [22] predictions are available, although the analogy between yielding and depinning remains controversial [2,23]. In these cases, avalanches are quantified from global variables (size, energy, duration,..). However, the shape, i.e. the internal temporal structure of the avalanche gives as well insightful information about the physical processes involved[24,25]. Universality is also expected in this case, meaning that the properly rescaled average avalanche velocity history $\langle v(t) \rangle$ should be shared by systems belonging to the same universality class [1]. For mean-field depinning, the avalanche shape *at fixed duration T* is parabolic, hence symmetric, $\langle v(t) \rangle_T \sim Tx(1-x)$ with $x=t/T$ [24], while *at fixed avalanche size* $S = \int_T v(t)dt$, the shape is of the form $\langle v(t) \rangle_S = S^{1/2} A \, y \, e^{-Ay^2/2}$ with $y = t/S^{1/2}$ and $A$ a non-universal constant [18,26]. $\tau = \left( \frac{2S}{A} \right)^{1/2}$ can be interpreted as a size-dependent characteristic time scale of the avalanche decay, with the above expression rewritten as $\langle v(t) \rangle_S = A\, t\, e^{-\left(\frac{t}{\tau}\right)^2}$, while the velocity is maximum at $t_M = \tau/\sqrt{2}$.

While this avalanche shape problem has been thoroughly analyzed theoretically [24,25,27,28], including beyond mean-field, relevant in case of short-range elasticity [26], experimental characterizations are still relatively scarce, likely because of the difficulty to track precisely the internal dynamics of individual events. Magnetic avalanches have been the most studied, and found to fit well the theoretical predictions of depinning [24,29]. In case of plastic avalanches in microcrystals, the generic form $\langle v(t) \rangle_S = A\, t\, e^{-Bt^C}$ was found for the shape at fixed size, though with a material-dependent exponent $C<2$, i.e. non universal [30]. In fracture and friction related problems, it has been argued that the seismic moment rate during earthquakes follows mean-field depinning predictions [31,32], hence supporting a model of an heterogeneous fault [18,33]. However, the data reported in [31] actually show an increasing leftward *asymmetry* of the shape at fixed *T* for larger avalanches, which accelerate faster than they decelerate [34]. A similar trend was observed for the intermittent propagation of stable cracks in mode I [14] and slip avalanches within granular media [5,6].

Here we focus on the progressive damage taking place under compressive loading within heterogeneous quasi-brittle materials such as rocks or concrete, which ultimately fail through the formation of an incipient fault. In that case, unlike earthquakes, the rupture does not take place along a pre-existing fault, but from a progressive localization of damage [35-39]. Unlike mode I fracturing, it involves friction. We previously mapped this problem onto the depinning of a *d*=3+1 interface represented by the damage field within the material, hence deduced predictions for (finite) size effects on strength that we found remarkably accurate for various rocks and concrete [40,41]. In this framework, and considering a local Coulomb's criterion for the onset of damage under compression [42-44], the equation of evolution of the damage field $D(\mathbf{r},t)$ reads [40]:

$$\mathrm{M}\frac{dD}{dt}(\mathbf{r},t) = R\left[ \sigma_C^{ext}(t) + \sigma_C^{el}(D,\mathbf{r}) - \sigma_C^{dis}(\mathbf{r}) \right] \quad (1)$$

, where M is a mobility coefficient (in Pa.s), $\mathbf{r}$ =(x,y,z) the position, $\sigma_C^{ext}$ the externally applied Coulomb stress, $\sigma_C^{el}$ the elastic contribution resulting from the damage field, $\sigma_C^{dis}$ the disorder accounting for material heterogeneities, and $R$ denotes the positive part. Indeed, at the time scale of the experiments (~s to ~minutes), damage is assumed to increase monotonically, i.e. healing processes are irrelevant. Eq. (1) assumes that the damage rate depends linearly on the excess of Coulomb stress compared with the local strength $\sigma_C^{dis}$.

Considering an elastic interaction kernel decaying with distance $r$ as $1/r^\alpha$, one could expect that such a $d$=3+1 interface would follow mean-field depinning for $\alpha \leq 3$. The Eshelby solution for the mechanical far-field around a damaged inclusion in an infinite 3D medium decays as $1/r^3$ [45], hence suggesting such mean-field scenario. However, a strong difference with classical depinning is the nonconvex nature of the kernel [13,40], allowing localization of damage along a fault [37]. Recently, tracking progressive damage within three types of concrete under uniaxial compression from acoustic emission (AE), we confirmed that compressive failure can be considered as a critical transition from an intact to a failed state, and found a surprisingly close agreement in terms of scaling exponents with mean-field depinning [13]. In that framework, the critical point, corresponding to the peak stress just preceding macroscopic faulting, is approached from below, i.e. $\Delta = \frac{\sigma_f - \sigma}{\sigma_f} \to 0^+$, where $\sigma_f$ is the failure stress and $\Delta$ the normalized distance from the critical point. In that study, however, we analyzed damage avalanches from their global AE signatures, energy and duration, without paying attention to their internal structure. Here, we explore the shape of these avalanches from a detailed analysis of the AE waveforms.

We recorded microfracturing-induced AE during the uniaxial compressive loading of cylindrical samples of three different concretes fabricated from different aggregate mixtures (fine *F*, i.e. only sand, medium *M*, and coarse C), and of four different sizes (*L*=80, 140, 220, 320 mm). All the details about the microstructural characteristics of the materials, the loading and AE recording protocol, have been given elsewhere [13,41]. Compression was applied at a constant stress rate of 0.5 MPa/s up to macroscopic failure. Two to four AE sensors with a frequency bandwidth of 20-1200 kHz and a resonant frequency of 900 kHz were fixed on each sample and their signals preamplified at 40 dB. AE bursts, signature of microcraking/damage avalanches, were detected over a 30 dB ($V_{th} \approx 3.2 \times 10^{-3}$ V) voltage amplitude threshold using a standard procedure [SM], and their waveforms saved at a sampling rate of 4 MHz (Fig. 1a). The wave intensity, or acoustic energy flux, is defined as $I(t)=V^2(t)$ (Fig. 1b) and the AE burst energy is integrated as $E_{AE} = \int_T I(t)dt$ over the waveform duration $T$.

The temporal evolution of the recorded intensity $I(t)$ is the convolution of a source signal $s(t)$ and a transfer function $G(t)$ describing the effects of the medium on wave propagation, $I(t)=(s*G)(t)$. In strongly heterogeneous materials like concrete, two distinct processes attenuate the source signal: (i) intrinsic absorption due to viscous and frictional effects at the microscopic scale, and (ii) scattering of waves by heterogeneities (sand particles, aggregates, microfracture gaps) [46,47]. The wavelength associated with the resonant frequency of the sensors, $f_R = 900$ kHz, is $\lambda_P = V_P / f_R \approx 5$ mm, where the P-wave velocity is $V_P \approx 4300$ m/s [41]. If this wavelength is smaller than the size of the scatterers and of the scattering mean free path, a multiple scattering regime might take place, for which the wave can interact several times with heterogeneities before reaching the sensor [47]. Hence, this mechanism is strongly unlikely for *F*-concrete characterized by a microstructural correlation length of 600 μm [SM]. For *C*- and *M*-concrete, with a microstructural correlation length of respectively 3.5 and 2.1 mm, it cannot be completely discarded a priori. If regime takes place, the transfer function $G(t)$ for a 3D medium would write:

$$G(t) = \frac{1}{(4\pi D_{iff} t)^{3/2}} e^{\frac{-r^2}{4 D_{iff} t}} e^{\frac{-t}{\tau_a}} \quad (2)$$

, where $r$ is the distance source-transducer, $\tau_a$ the absorption time scale, and $D_{iff}$ the diffusivity. In concrete, $D_{iff}$ depends on the aggregate content but is of the order of 20 mm²/μs at a frequency of 100 kHz, and 10 mm²/μs at 1 MHz [47]. In cases where multiple scattering can be neglected, the transfer function (2) summarizes to an exponential decay $G(t) = e^{\frac{-t}{\tau_a}}$, similar to the absorption model proposed by [48]. As a matter of fact, all the evidences argue against a significant role of multiple scattering on our results, whatever the material considered. Indeed, a signature of a power law decay $t^{-3/2}$ is not detectable in our avalanche profiles, including at short timescales. More importantly, the scattering mean free path and the diffusivity depend on the type of concrete, while it is shown below that our results do not. In principle, one could from (2) deconvoluate the recorded intensity to obtain the source signal. However, as the AE sources were not localized, we followed a different path: we show that, for long enough signals ($T \gg 100$ μs), the shape of the source intensity $s(t)$ is only marginally modified by $G(t)$. Hence, the recorded signal $I(t)$ retains the main characteristics of $s(t)$, and particularly its asymmetry, which cannot be explained by multiple scattering or absorption [SM]. We also checked that the silicone gluing of the sensors cannot explain this asymmetry, as it does not influence the shape of the convoluted signal at timescales larger than the hit definition time (HDT=20 μs) determined from pencil lead break tests [SM].

Figs. 1c and 1d show respectively the relationship between the conditional average maximum intensity $\langle I_{max}|T\rangle$, the conditional average AE energy $\langle E_{AE}|T\rangle$, and the duration of the AE events. Assuming a pulse-like event at the source with a short rise time, then modified by intrinsic absorption during wave propagation within the medium with an absorption time scale $\tau_a$, one would expect $\langle I_{max}|T\rangle = V_{th}^2 \exp(T/\tau_a)$, as well as $\langle E_{AE}|T\rangle = (\tau_a/2)V_{th}^2(\exp(2T/\tau_a)-1)$. This behavior is recovered for short durations, using $\tau_a \sim 100\mu s$ (Figs. 1c and 1d), hence signing an exponential decay of $I(t)$ resulting from intrinsic absorption. Consequently, for short, pulse-like events, the AE duration is meaningless in terms of source mechanisms, i.e. the internal structure of these events cannot be studied from their AE signature. On the reverse, for longer avalanches $(T \gg \tau_a)$, non-trivial scaling are observed, $\langle I_{max}|T\rangle \sim T^\delta$, with δ=2.0±0.3, as well as $\langle E_{AE}|T\rangle \sim T^k$ with $k$=2.2±0.2 (Fig. 1d). In this regime, we used a previously detailed elastic crack/fault model to relate the AE characteristics to the source mechanism [13]. This source model, relevant for both mode I cracks and shear faults, assumes an average slip or displacement $\langle u \rangle$ proportional to the crack or fault "radius" $\langle u \rangle \sim A^{1/2}$, where $A$ is a compact (nonfractal) incremental crack or fault area, an $A$-independent stress drop, and a constant scaled energy, i.e., a radiated acoustic or seismic energy $E_{AE}$ proportional to the potency $P_0 = <u>A$ representing the size $S$ of the avalanche. This model is supported experimentally [49,50], although this issue is not yet entirely solved: such $E_{AE} \sim P_0$ scaling clearly emerges when combining several datasets (experimental and field studies) covering a wide range of scales, but individual catalogs might be better described using larger exponents, $E_{AE} \sim P_0^g$, with $g$>1, possibly as a consequence of limited frequency bandwidths of the sensors [50]. Considering a scaling $E_{AE} \sim P_0$, the AE intensity $I(t) = dE_{AE}/dt$ is a measure of the potency rate, i.e. of the avalanche "velocity" $v$. Hence, the avalanche shape $v(t)$ can be retrieved over the avalanche duration from $I(t)$, low-pass filtered at $f_c = 1/\tau_a \sim 10$ kHz. As the average avalanche shapes discussed below were obtained from the stacking of many uncorrelated individual signals, the results were found to be independent of this low-pass filtering [SM].

The scaling between the global variables, AE energy (i.e. avalanche size) and duration (Fig. 1d), is compatible with mean-field depinning that predicts $k$ =2 [21,22], hence with our previous mapping [13,40]. The analysis of the internal avalanche dynamics reveals, however, a different story. Figure 2 shows the scaled avalanche shapes at fixed duration, $\langle v(t) \rangle_T / T$. A clear departure from mean-field predictions is observed as (i) the shapes show a strong leftward asymmetry and (ii) do not collapse after the rescaling by *T*. We quantified this asymmetry from the skewness $\zeta$ of the shape [34], which is large ($\zeta \approx 1.3$) and independent of *T* for avalanche durations larger than 200 µs. All these results are found to be essentially independent of the distance to the critical point $\Delta$, of the sample size *L* as well as the concrete mix (*F*, *M* or *C*) (Fig. 2a and [SM]). While at odds with mean-field depinning, this asymmetry is reminiscent of, but more pronounced than observations for earthquakes [31,34], mode I crack propagation increments [14], or for 2D sheared granular media [5]. Leftward asymmetry has been also reported for avalanches in domain walls dynamics, though much slighter than in our case ($\zeta \leq 0.1$) [34,51]. This remarkable asymmetry also expresses through a slow, power-law decay, $\langle v(t) \rangle_T / T \sim t^{-p}$ with $p \approx 2.05$, for $t/T \rightarrow 1$ (Fig. 2c).

The shapes at fixed size *S* (i.e. at fixed AE energy $E_{AE}$) do not follow mean-field depinning predictions as well (Fig. 3). For small avalanches, an exponential decay is observed (Fig. 3a), reminiscent of the intrinsic absorption discussed above. Larger avalanches decay much more slowly. Following previous work [30], we fitted the shapes using the form $\langle v(t) \rangle_S = A t e^{-Bt^C}$. This expression can be rewritten as $\langle v(t) \rangle_S = A t e^{-\left(\frac{t}{\tau}\right)^C}$, with the time constant $\tau = B^{-1/C}$. Note that a constant initial acceleration, i.e. $\langle v(t) \rangle_S \sim t$ at small *t*, is consistent with *k* close to 2 [22]. We obtain a *stretched* exponential deceleration, with $C=0.60 \pm 0.18$ for avalanches recorded over the entire loading path ($1 \geq \Delta \geq 0$) and $C=0.48 \pm 0.17$ for avalanches close to the peak stress ($\Delta \leq 0.1$), i.e. a value at odds with the mean-field depinning prediction (*C*=2). The time constant $\tau$ is in the range 10-20 µs. As for the avalanches at fixed duration, we did not find a significant dependence of these parameters on the distance to the critical point $\Delta$, the sample size *L*, or the concrete mix (*F*, *M* or *C*) (Fig. 3b and [SM]). This stretched deceleration cannot result from the transfer function $G(t)$, which sole effect would be to slightly *increase C*, depending on the distance source-sensor and the diffusivity [SM]. The independence of the avalanche shape on the concrete mix, i.e. on scattering mean free path and diffusivity, argues as well for a marginal effect of wave scattering. Similarly, the fact that avalanche shapes do not evolve as approaching failure ($\Delta \rightarrow 0$) argues for a negligible impact of microfracture gaps. Hence, this slow deceleration is a genuine characteristic of the source mechanism and expresses the asymmetry of the avalanche shape at fixed avalanche size. As shown on Fig. 3b, this stretched exponential form actually overestimates the deceleration at large times, i.e. the damage avalanche is decaying even more slowly before reaching the background noise. Indeed, we show on Fig. 3c that the decelerating part of the avalanche shapes at fixed size exhibits, for large avalanches, a power-law tail, $\langle v(t) \rangle_S \sim t^{-p}$ for large $t$ with $p \approx 2.1$, reminiscent of the shapes at fixed duration (Fig. 2c). Overall, although our data lie in between the stretched exponential form and the power-law decay, the leftward asymmetry is remarkable.

To explain this asymmetry, the *quasi*-brittleness of concrete can be put forward. Quasi-brittle materials are characterized by the development of a fracture process zone (FPZ) ahead of a propagating fracture or fault [52]. Within the FPZ, enhanced material softening and viscoelastic mechanisms take place. In concrete, these non-brittle processes of energy dissipation taking place within the cement paste can have various microscopic origins, which remain partly obscure [53]. The rearrangement of nanoscale particles of calcium-silicate-hydrates (C-S-H), which are the primary component of the cement paste, might play a major role [53]. At much larger length (entire sample size) and time (hours to days) scales, this viscoelasticity has been modelled using Kelvin-Voight rheologies [52,54], and coupled to damage mechanics [55]. At microscopic length scales and under the stress concentrations at the tip of a fracture or fault, significant viscoelasticity will take place over much shorter timescales [53]. It generates delayed inelastic strains, hence retardation effects that might explain the asymmetry of the damage avalanche shape [26], as already argued for Barkhausen noise [34,51]. In this last case, retardation effects, which are much less pronounced than in the present case, come from energy dissipation by eddy currents and can be modelled by the introduction of a damping memory kernel in the equation of motion of the wall. Interestingly, viscosity in the particle displacements, induced by basal static friction, was invoked to explain avalanche shape asymmetry within a 2D sheared granular medium, although a retardation mechanism was not explicitly considered in this case [5]. In our case, viscoelastic softening and delayed stress relaxation locally occurs within the FPZ generated by the damage avalanche as it progresses, inducing a negative feedback on this propagation. The evolution equation (1), which only takes into account elasticity, purely brittle damage, and disorder, cannot account for these effects. Building an analogy with [26], we can introduce a retardation kernel $f$, non-local in time, within (1):

$$M \frac{dD}{dt}(\mathbf{r},t) + a \int^t f(t-t') \frac{dD}{dt}(\mathbf{r},t') dt' = R \left[ \sigma_C^{ext}(t) + \sigma_C^{el}(D,\mathbf{r}) - \sigma_C^{dis}(\mathbf{r}) \right] \quad (3)$$

, where $a$ is a coefficient setting the strength of the retardation effect and having the same units as M, i.e. that of a viscosity (Pa.s), while $f(x=0)=1$ and then monotonously decreases towards 0 as $x \to +\infty$. This introduces an asymmetry of the avalanche shape whose magnitude depends on the kernel $f$ and the ratio $a/M$ [26]. With this scenario, the avalanche shape is not expected to show necessarily universal features. The tail of the shape function is likely a fingerprint of the retardation kernel. Indeed, viscoelastic memory kernels have been introduced theoretically in a generic depinning framework, though, for simplicity, with a Maxwell-like exponential decay, $f(x) = e^{-t/\tau}$ [56], and it was shown that the tail of the avalanche shape at fixed size reproduces this form of the memory kernel, with the same time constant τ [26]. As explained above, the exponential decay of the shape function observed in our case for small durations most likely results from the absorption of AE waves, and is not representative of the source mechanism. For avalanches of duration $T \gg \tau_a \approx 100\mu s$, we can guess that our memory kernel $f(x)$ should be characterized by a slower decay, in between a stretched exponential and a power law decay. Hence, our results argue for non-Maxwell viscoelastic processes within the FPZ, which are different from "bulk" wave absorption taking place within the undamaged material.

Such retardation effects induced by localized viscoelasticity within the cement paste, i.e. independently of the aggregate content, are consistent with a damage avalanche shape independent of the concrete mix [SM]. It has been shown that such retardation mechanisms modify the avalanche shape but does not affect the distributions of avalanche sizes or durations [26,51]. Consequently, the present results, while strongly departing from mean-field depinning predictions in terms of avalanche shape, remain compatible with our previous mapping based on distributions and scaling relationships between global variables [13,40]. In other words, viscoelasticity within the FPZ affects the damage dynamics at short time scales only (few ms at most). At larger timescales, the dynamics remains essentially controlled by the applied stress, internal disorder and elastic stress redistributions.

In their theoretical analysis of a modified ABBM model [57] including retardation effects, Dobrinevski et al. [26] showed that the avalanche shape asymmetry is associated with a "breakup" of avalanches into sub-avalanches, or "aftershocks". Such sub-avalanches are not detectable on the average shape profiles of Fig. 3, but can be easily identified on various individual avalanche profiles (examples are given on Fig. 4). Interestingly, in these examples, the asymmetry of the dynamics is clearly recognizable for each sub-avalanche. Such time clustering of sub-avalanches is also reminiscent of the relation between post-seismic relaxation and aftershock triggering along natural faults[58]. Available analyses argue for a weaker asymmetry of earthquake *co*-seismic ruptures compared with our results [31]. However, the very slow decay of damage avalanches reported here is reminiscent of the *post-seismic* slip rate decay detected after large earthquakes from geodetic measurements (and not from detectable elastic waves as in our case) [58,59]. Aseismic afterslip rates decay in $t^{-p}$ with $p$ close to 1, much like the associated Omori's decay [60] of aftershock triggering. Interestingly, transient viscoelastic effects within a so-called brittle creep fault zone, downdip the co-seismic rupture and equivalent to our FPZ, were put forward to explain such slow decay of post-seismic slip [61]. Whether this analogy is a signature of a common mechanism, spanning a huge range of space and time scales, remains an open question.

## ACKNOWLEDGEMENTS

This work has been supported by the AGIR program of Université Grenoble-Alpes. We thank Jean Virieux and Philippe Roux for highly valuable discussions on diffusive scattering, Philippe Le Doussal for discussions about retardation effects, as well as two anonymous referees for highly valuable comments on the manuscript.

* jerome.weiss@univ-grenoble-alpes.fr

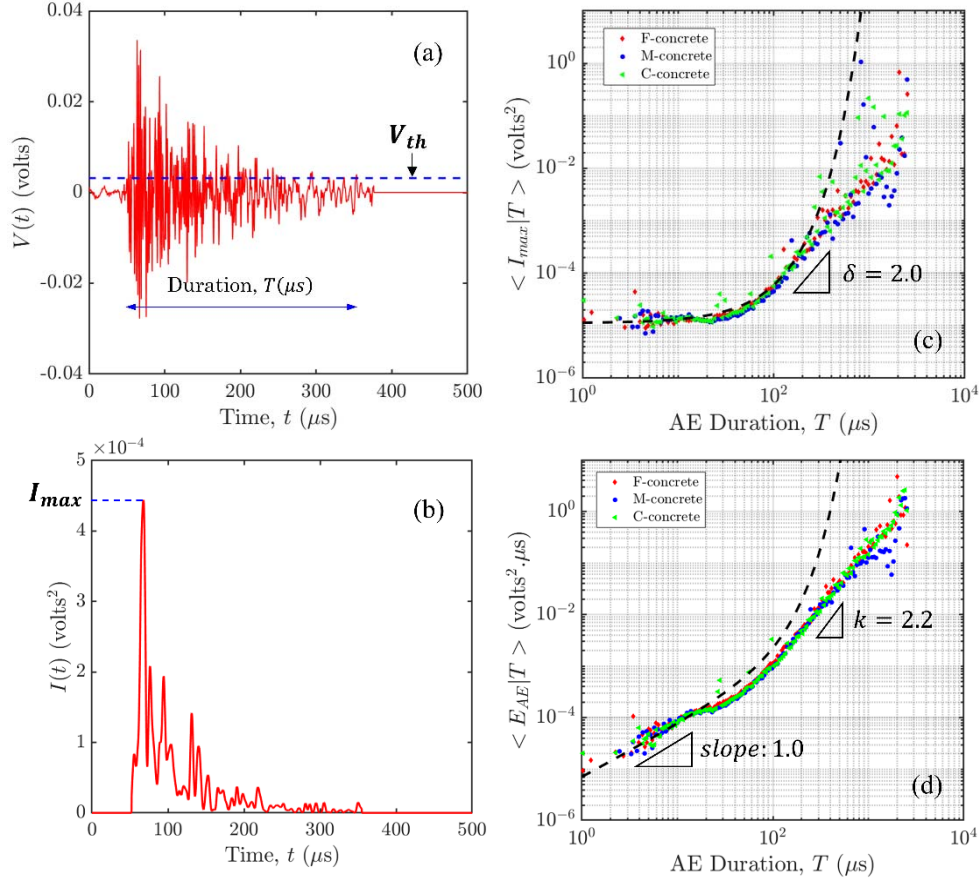

FIG. 1. (a) Example of an acoustic waveform recorded during compression of a sample of C-concrete. (b) Corresponding intensity record, $I(t) = V^2(t)$. (c) Relationship between the conditional average maximum intensity $\langle I_{max} | T \rangle$ and the duration of the AE events, $T$. The dotted black line represents the equation $\langle I_{max} | T \rangle = V_{th}^2 \exp(T/\tau_a)$, with $\tau_a = 100$ μs. (d) Relationship between the conditional average AE energy $\langle E_{AE} | T \rangle$ and the duration. The dotted black line represents the equation $\langle E_{AE} | T \rangle = (\tau_a / 2) V_{th}^2 (\exp(2T/\tau_a) - 1)$ expected for a pulse-like event at the source modified by intrinsic absorption during wave propagation. For each concrete material, this is obtained from a compilation of data coming from the four different sample sizes, the different AE sensors, and the events recorded during the entire loading ($0 \leq \Delta \leq 1$).

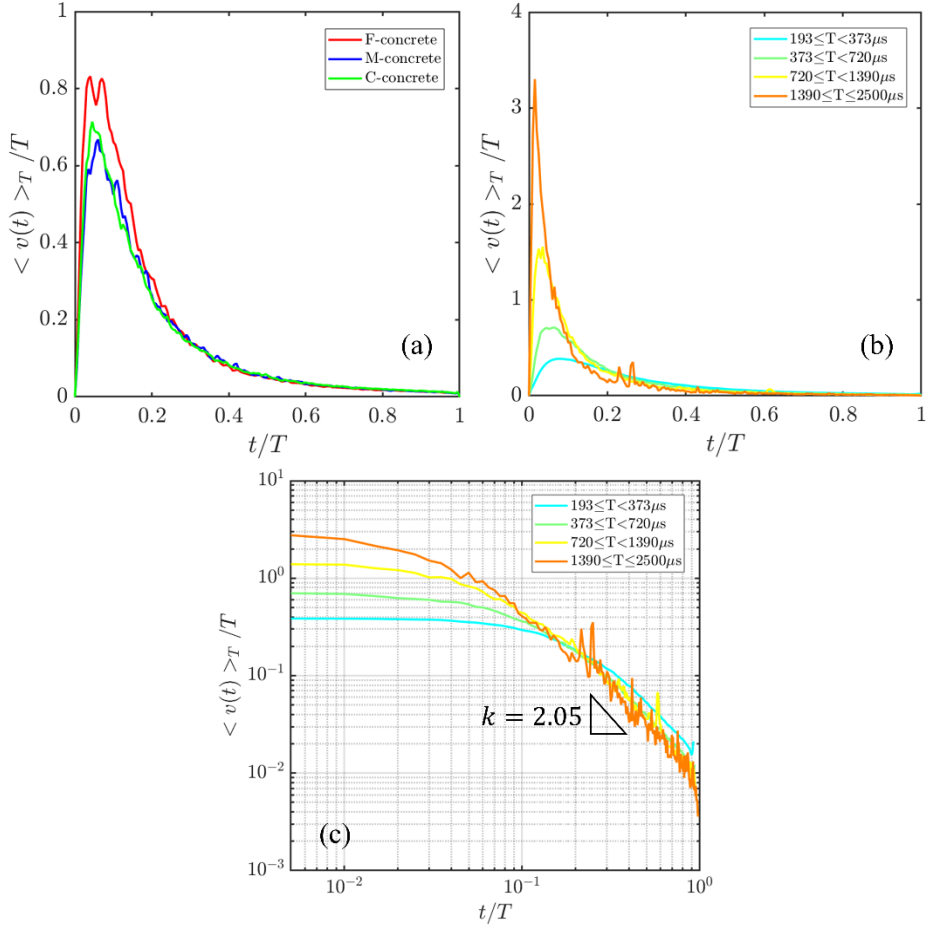

FIG. 2. Scaled avalanche shapes at fixed duration, $\langle v(t) \rangle_T / T$. (a) Scaled shapes for avalanches of durations 373≤$T$≤720 μs recorded during the entire loading (0≤Δ≤1), for the three concretes. For each concrete, averaging is performed over AE waveforms coming from the four different sample sizes, and recorded by the different AE sensors. (b) Scaled shapes for avalanches of different durations, recorded during the entire loading (0≤Δ≤1). Averaging is performed over AE waveforms coming from the three concrete materials, the four different sample sizes, and recorded by the different AE sensors. (c) Decaying parts ($t>t_M$) of the scaled shapes of (b) shown in a log-log representation.

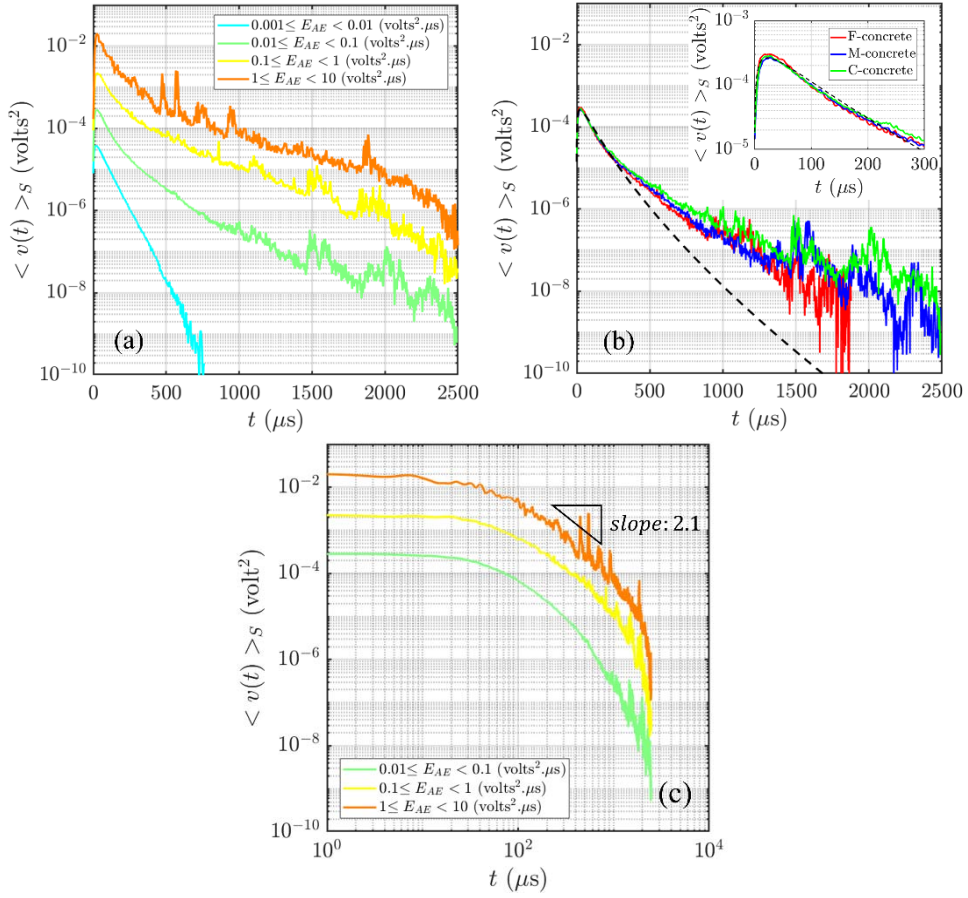

FIG. 3. Avalanche shapes at fixed size, $\langle v(t) \rangle_S$ : Stretched exponential form. (a) Shapes for avalanches of different sizes, recorded during the entire loading ($0 \leq \Delta \leq 1$). Averaging is performed over AE waveforms coming from the three concrete materials, the four different sample sizes, and recorded by the different AE sensors. The exponential decay $e^{\frac{-t}{\tau_a}}$ observed for $10^{-3} \leq S \leq 10^{-2}$, with $\tau_a$=65 µs, is the fingerprint of intrinsic absorption of acoustic waves. (b) Shapes for avalanches of size $0.01 \leq S \leq 0.1$ recorded during the entire loading ($0 \leq \Delta \leq 1$), for the three concretes. For each concrete, averaging is performed over AE waveforms coming from the four different sample sizes, and recorded by the different AE sensors. The black dotted line represents the equation $\langle v(t) \rangle_S = A t e^{-Bt^C}$, with $A=4\times10^{-4}$, $B$=0.20 and $C$=0.64. The inset is a focus of the main panel over the first 300 µs. (c) Decaying parts ($t > t_M$) of the shapes of (a) shown in a log-log representation.

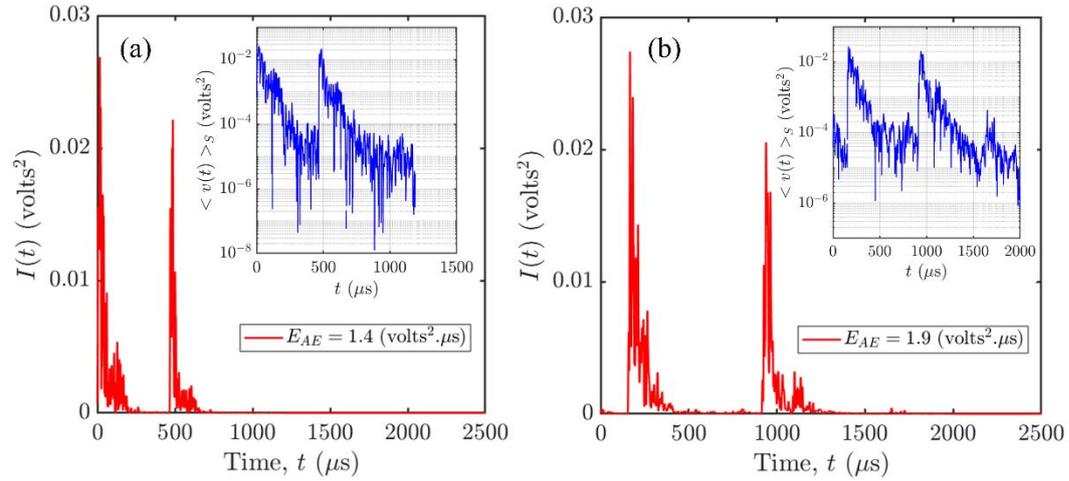

FIG. 4. Breakup of avalanches into sub-avalanches, or "aftershocks". (a) An example of an individual avalanche profile recorded in F-concrete. The inset represents the same profile in a semi-log representation. (b) Idem for C-concrete.

# Asymmetric damage avalanche shape in quasi-brittle materials and sub-avalanches (aftershocks) clusters

-

# Supplemental Material

Chi-Cong Vu[1] and Jérôme Weiss[2,*]

[1]*National University of Civil Engineering, Ha Noi, Viet Nam*
[2]*Univ. Grenoble Alpes, CNRS, ISTerre, 38000 Grenoble, France*

I. Concrete samples and acoustic emission recording

The preparation of the concrete samples as well as the mechanical experimental setup have been extensively described elsewhere [40]. The three concretes (Fine, Medium and Coarse) were prepared from different aggregate size distributions. The corresponding aggregate grading curves are shown on Fig. S1, giving median aggregate size of respectively 500 μm, 7 mm and 15 mm.

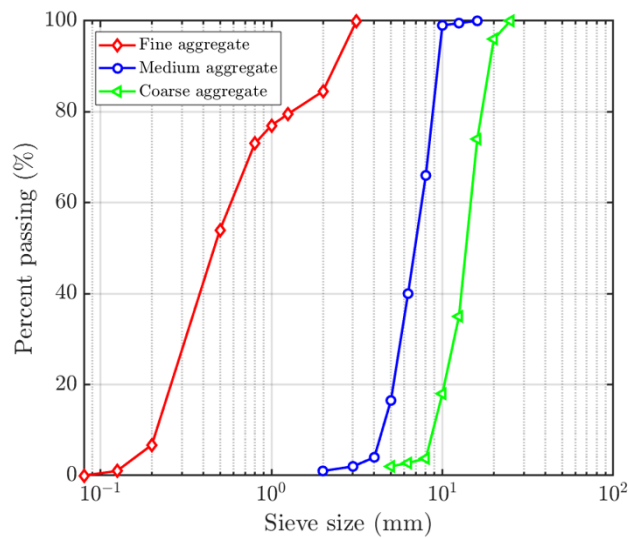

Fig. S1. Grading curves of aggregates for the three concrete mixtures.

The microstructural characteristics of the three concretes analyzed were obtained from image analyses of internal sections such as those of Fig. S2 [40]. They are summarized in Table S1. The correlation length of the microstructure expresses both the average aggregate size and the average distance between these aggregates, i.e. the scattering mean free path for acoustic waves. This correlation length is of the order of the median aggregate size.

| Concrete Mixture | Correlation length of the global microstructure (mm) | Correlation length of the pore microstructure (μm) | Mean pore diameter (mm) | Maximum pore diameter (mm) | Porosity |
|---|---|---|---|---|---|
| Fine | 0.6 | 26.4 | 0.33 | 6.9 | 0.048 |
| Medium | 2.1 | 9.8 | 0.31 | 6.7 | 0.016 |
| Coarse | 3.5 | 8.5 | 0.28 | 5.4 | 0.015 |

Table S1. Main microstructural characteristics of the three concrete materials.

For each concrete, Acoustic Emission (AE) was recorded during compression tests performed on four samples sizes: 4 tests for $L$=80 mm samples, and 2 tests for $L$=140, 220 and 320 mm samples. This represented a total of 30 mechanical tests with AE recording. The avalanche shapes presented in the main text were obtained from an averaging over the different samples of a given material. Indeed, we checked that these avalanche shapes are independent of the sample size (see section IV below).

AE sensors of type PICO produced by Physical Acoustics Corporation were used. Their frequency range is about 20 kHz-1.2 MHz, with a peak frequency of approximately 900 kHz. Their small size (4×5 mm) make them easy to couple on our small samples (i.e. $L$=80-mm samples). We used a Silicone adhesive glue (Silcoset 151) for coupling. In order to ensure a proper coupling, some small areas on the lateral surfaces of the samples were ground and polished by an angle grinder with less and less grit size of metal-bonded discs. This ensured that the sample surfaces were not damaged. The AE signals from the loaded specimen are converted into electrical signals by the AE sensors, then pre-amplified with a gain of 40 dB and recorded by the Acoustic Emission Digital Signal Processor (AEDSP-32/16) cards at a sampling rate of 4 MHz. To detect AE bursts, we used an amplitude threshold of 30 dB, a Peak Definition Time (PDT) of 10 μs, a Hit Definition Time (HDT) of 20 μs and a Hit Lock-out Time (HLT) of 20 μs. These parameters were defined by performing AE recording on samples before loading but with the loading machine switched on (to set the AE threshold relatively to the environmental noise amplitude), as well as Pencil Lead Break (PLB) tests which are similar to Hsu-Nielsen tests (to set PDT, HDT and HLT). These PLB tests were performed on the samples surface before mechanical loading, close to the sensor to minimize the effect of wave travelling within the concrete. These timescales, and particularly the HDT, express the typical timescale of the response of the Silicone glue and the sensor to a very short, pulse-like source. Consequently, the convolution of a source signal with the silicone gluing and sensor transfer function will not significantly affect the shape of the recorded signals at timescales larger than 20 μs.

The burst duration $T$ is defined as the time over which the envelope of the AE signal $V(t)$ remains above the threshold $V_{th}$=30 dB.

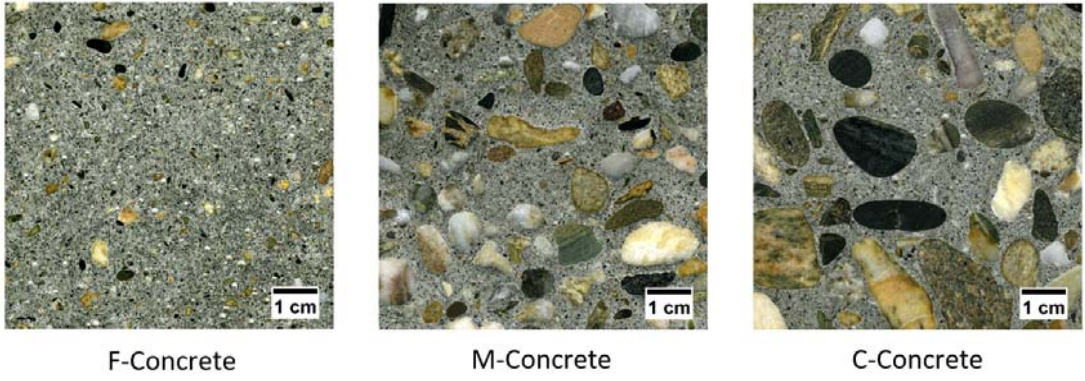

Fig. S2. Cross sections of the three different concrete mixtures used for the experimental investigations

II. Effect of multiple scattering and intrinsic absorption on the recorded AE signals

We consider a source function (the avalanche shape) emitted at the source, $s(t)$. The intensity of the signal recorded at the sample surface by a sensor, $I(t)$, is a convolution of this source function with a Green's function $G(t)$ describing the effects of the medium on travelling waves, $I(t) = (G * s)(t)$. In strongly heterogeneous materials like concrete, an acoustic wave emitted at a source will be attenuated within the material before reaching the sensor, following two distinct physical phenomena: (1) scattering and (2) intrinsic absorption [46].

Intrinsic absorption is due to viscous effects. It is characterized by a damping time scale $\tau_a$, and writes:

$$G_{(abs)}(t) = e^{-t/\tau_a} \quad (1)$$

The scattering regime depends on the relative value between the AE wavelength, $\lambda$, the size of heterogeneities, $d$, and the scattering mean free path (average distance between scatterers), $l$. For P-waves, $\lambda_p = V_p/f$, with $V_p \approx 4300$ m/s for concrete [40]. Taking a resonant frequency of 900 kHz for the transducers, we get $\lambda_p \approx 5$ mm. Therefore, for F-concrete, $\lambda_p$ is one order of magnitude larger than the median aggregate size and the microstructural correlation length (see above). In this case, we do not expect multiple scattering to take place [46]. For M- and C-concrete, $\lambda_p$ is of the order of magnitude of these microstructural length scales, and so we cannot discard completely a possible role of multiple scattering [46], for which, in a 3D medium and assuming a constant and isotropic diffusivity, one has:

$$G_{(diff)}(t) = \frac{1}{(4\pi Dt)^{3/2}} e^{\frac{-r^2}{4D_{iff}t}} \quad (2)$$

, where $r$ is the distance source-transducer, and $D_{iff}$ the diffusivity. Near 1 MHz, $D_{iff}$ is in the range 5 to 10 mm²/μs in concrete [46]. The combination of (1) and (2) gives:

$$G(t) = \frac{1}{(4\pi Dt)^{3/2}} \times e^{\frac{-r^2}{4D_{iff}t}} \times e^{-t/\tau_a} \quad (3)$$

In principle, knowing the recorded signal $I(t)$, one could deconvoluate this signal using (3) to retrieve the source signal. However, the role of multiple scattering is uncertain (see above), and the avalanche shape is obtained from the averaging over many individual events, whose distances from the sensor are unknown. In what follows, we use another methodology: We check that the recorded avalanche shapes,

$$I(t) = A\, t\, e^{-(t/\tau)^C} \quad (4)$$

, with $C \ll 2$ (see main text), *cannot* result from the convolution of a simple symmetric source signal with the Green's function (3), and that a source function of the form of (4) would be only marginally modified by multiple scattering and intrinsic absorption.

If the absorption time scale $\tau_a$ is small, and the source signal is pulse-like, the term $e^{-t/\tau_a}$ will dominate. This is the case for e.g. a Dirac signal of duration $T < \tau_a$. In this case, $I(t) \approx I_{max} e^{-t/\tau_a}$, and so one should have the following relation between $I_{max}$ and duration $T$:

$$I_{max} \approx V_{th}^2 = cte, \text{ for } T < \tau_a$$

This is the regime we observe at small durations, see Fig. 1c of the main text. In this regime, the intensity $I(t)$ is expected to exponentially decay with time, as observed (see Fig. 3a of the main text for avalanche of AE energy below 0.01 V².µs). Similarly, the conditional average acoustic energy $\langle E_{AE}|T\rangle$ of the convoluted signal is expected to scale as $\langle E_{AE}|T\rangle = (\tau_{a/2})V_{th}^2(e^{2T/\tau_a} - 1)$. At very short durations ($T \ll \tau_a$), this would mean $\langle E_{AE}|T\rangle \sim T$, as observed (Fig. 1d).

We consider now a *symmetric* source signal $s(t)$ of duration $T \gg \tau_a$, and the impact of the convolution of $s(t)$ with the Green's function (3). The parameters of the Green's function are:

- Diffusivity $D_{iff}$: we take 10 mm²/µs [46]
- Absorption time scale $\tau_a$
- Distance source-transducer $r$
- Duration of the source signal $T$

Considering e.g. a triangular source signal $s(t)$ of duration $T = 500$ µs, $r=50$ mm and $\tau_a = 100$ µs (see main text), the resulting convoluted signal $I(t) = (G * s)(t)$ is shown on Fig. S3. The triangular shape is shifted towards larger times due to an initial delay, but the asymmetry induced by the attenuation remains small. If one tries to fit this convoluted shape with the generic expression (4), one finds a $C$ exponent larger than 5 but the fit is not very good.

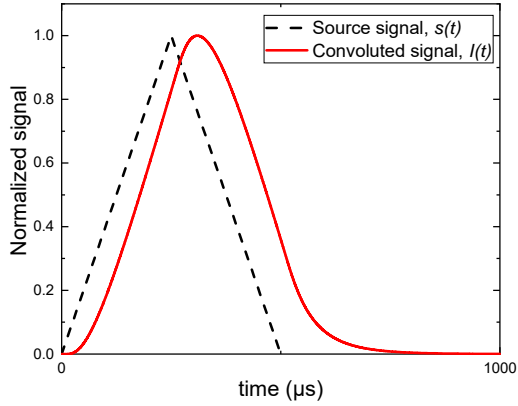

Fig. S3. Triangular source signal $s(t)$ (dashed black line), and the resulting intensity $I(t)$ (red solid line) after convolution with the Green's function (3).

Considering now a source signal with a mean-field depinning shape, $s(t) \sim t e^{-\left(\frac{t}{\tau}\right)^2}$ with $\tau$=300 µs, while keeping the parameters of the Green's function unchanged, the resulting recorded signal $I(t)$ is shown on Fig. S4. We observe again a time-shift, but the overall shape remains similar. If we best-fit this convoluted signal $I(t)$ with the generic form (5), we find $C$=3.1±0.0004 and $\tau$=448 µs. Hence, the effect of wave diffusion and absorption is to increase the time scale $\tau$, and to *increase* the exponent $C$. This, therefore, cannot explain the observed exponents $C < 2$.

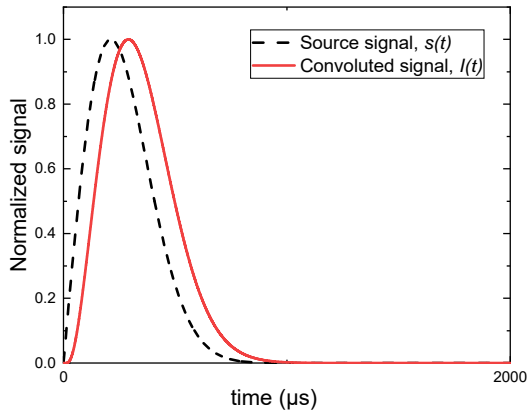

Fig. S4. Mean-field depinning source signal $s(t) \sim t e^{-\left(\frac{t}{\tau}\right)^2}$ (dashed black line), and the resulting intensity $I(t)$ (red solid line) after convolution with the Green's function (3).

We finally consider a source signal $s(t)$ of the form (4) with, following our results detailed in the main text, $C = 0.5$ and $\tau = 10$ µs. Keeping the parameters of the Green's function unchanged, the resulting recorded signal $I(t)$ is shown on Fig. S5. If we best-fit this convoluted signal with the generic form (4), we obtain $C$ =1.32±0.0003 and $\tau$=181 µs.

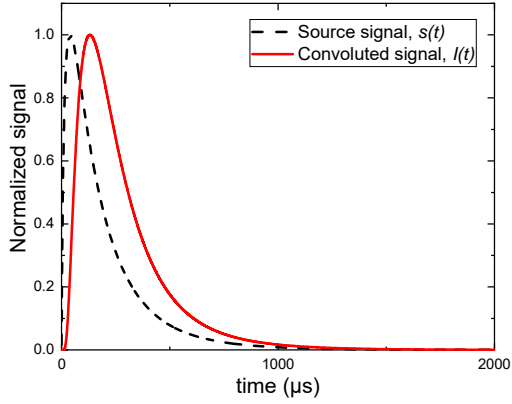

Fig. S5. Source signal $s(t) \sim t e^{-\left(\frac{t}{\tau}\right)^C}$, with $C$=0.5 and $\tau$=10 μs (dashed black line), and the resulting intensity $I(t)$ (red solid line) after convolution with the Green's function (3).

Hence, the effect of wave diffusion and absorption is to *decrease* the asymmetry: the avalanche appears to start more slowly, and to decelerate slightly more rapidly. This effect depends on the distance source-transducer: if we now consider a source at $r$=10 mm from the transducer, while keeping $D_{iff}$ =10 mm²/μs and $\tau_a$ =100 μs, we find $C$=0.66±0.0001 and $\tau$=33 μs. It depends also on the diffusivity: For $D$ =30 mm²/μs, $r$=50 mm and $\tau_a$ =100 μs, one finds $C$=0.97±0.0001 and $\tau$=100 μs.

From this, if the source signals were significantly modified by wave scattering, one would expect:

(i) As a larger distance $r$ means also a smaller recorded maximum intensity, one would expect slightly larger $C$-values for (apparently) *smaller* avalanches. This is not observed.

(ii) The diffusivity is expected to change with the size of the scatterers, so with the concrete mix. This should impact the apparent shape if the effect of wave scattering is strong. We do not observe any significant material dependency on the avalanche shape (see e.g. Fig. 3b of main text), hence arguing against an important role of wave scattering on the observed shapes.

III. The independence of the avalanche shapes on low-pass filtering

The avalanche shapes discussed in the main text have been obtained from the averaging of many intensity signals $I(t)$, low-pass filtered at $f_c = 10\ kHz$. However, as shown on Fig. S6 for C-concrete, as this averaging was performed over many uncorrelated individual signals, the results are independent of this low-pass filtering.

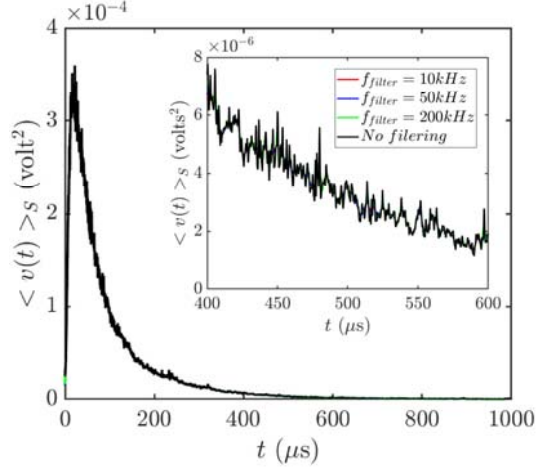

Fig. S6. Average avalanche shape for avalanches of sizes $0.01 \leq S \leq 0.1$ for C-concrete, using various low-pass filtering frequencies for the individual signals, or no filtering.

IV. The effects of sample size, and distance to the critical point on the average avalanche shape.

The avalanche shapes were found to be independent of (i) the concrete mix (see main text), (ii) the sample size (Fig. S7a), and (iii) the distance to the critical point $\Delta = \frac{\sigma_f - \sigma}{\sigma_f}$ (Fig. S7b).

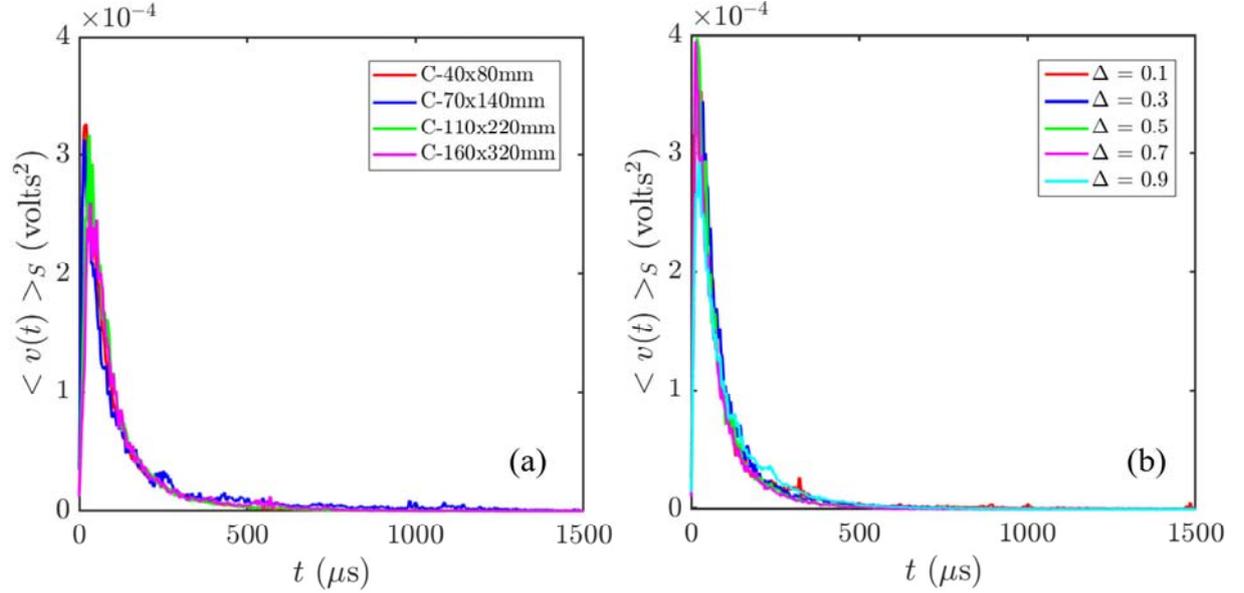

Fig. S7. (a) Average avalanche shape for avalanches of sizes $0.01 \leq S \leq 0.1$ for C-concrete samples of different sizes (diameter × length). The average is performed over AE events recorded during the entire loading history ($0 \leq \Delta \leq 1$). (b) Average avalanche shape for avalanches of sizes $0.01 \leq S \leq 0.1$ for C-concrete, for events recorded at different distances from the critical point, $\Delta$. For each $\Delta$-value, the average is performed over samples of different sizes (diameter × length). Similar results were obtained for other avalanche sizes.

The corresponding skewness values for events recorded during the entire loading and the three materials are giving on Table S2, while the values obtained for C-concrete at different distances from the critical point are shown on Table S3.

| Avalanche size | Sample size | F-Concrete | M-Concrete | C-Concrete |
|---|---|---|---|---|
| $0.001 \leq S < 0.01$ | 40x80mm | 1.32 | 1.32 | 1.32 |
|  | 70x140mm | 1.32 | 1.39 | 1.32 |
|  | 110x220mm | 1.32 | 1.32 | 1.32 |
|  | 160x320mm | 1.32 | 1.31 | 1.32 |
|  | All sample sizes | 1.32 | 1.33 | 1.32 |
| $0.01 \leq S < 0.1$ | 40x80mm | 1.32 | 1.33 | 1.33 |
|  | 70x140mm | 1.33 | 1.39 | 1.35 |
|  | 110x220mm | 1.33 | 1.32 | 1.33 |
|  | 160x320mm | 1.33 | 1.32 | 1.34 |
|  | All sample sizes | 1.33 | 1.33 | 1.33 |
| $0.1 \leq S < 1$ | 40x80mm | 1.33 | 1.34 | 1.34 |
|  | 70x140mm | 1.35 | 1.38 | 1.38 |
|  | 110x220mm | 1.35 | 1.31 | 1.34 |
|  | 160x320mm | 1.33 | 1.33 | 1.34 |
|  | All sample sizes | 1.34 | 1.34 | 1.34 |
| $1 \leq S < 10$ | 40x80mm | 1.34 | 1.33 | 1.33 |
|  | 70x140mm | 1.31 | 1.35 | 1.39 |
|  | 110x220mm | 1.35 | 1.33 | 1.32 |
|  | 160x320mm | 1.34 | 1.33 | 1.35 |
|  | All sample sizes | 1.33 | 1.34 | 1.34 |

Table S2. Skewness values of the avalanche shapes, averaged over the entire loading, for different sample sizes and different concrete mixtures

| Avalanche size | Skewness | | | | |
|---|---|---|---|---|---|
|  | $\Delta = 0.1$ | $\Delta = 0.3$ | $\Delta = 0.5$ | $\Delta = 0.7$ | $\Delta = 0.9$ |
| $0.001 \leq S < 0.01$ | 1.31 | 1.31 | 1.32 | 1.32 | 1.32 |
| $0.01 \leq S < 0.1$ | 1.34 | 1.33 | 1.32 | 1.32 | 1.33 |
| $0.1 \leq S < 1$ | 1.34 | 1.33 | 1.34 | 1.33 | 1.34 |
| $1 \leq S \leq 10$ | 1.34 | 1.37 | 1.33 | 1.32 | 1.33 |

Table S3. Skewness values of the avalanche shapes for different distances from the critical point for C-concrete samples